\documentclass[11pt, epsf]{article}
\usepackage{epsfig}
\newcommand{\be}{\begin{equation}}
\newcommand{\ee}{\end{equation}}
\newcommand{\bea}{\begin{eqnarray}}
\newcommand{\eea}{\end{eqnarray}}
\newcommand{\rp}{r_+}
\newcommand{\re}{r_{ex}}
\def\ga{\mathrel{\raise.3ex\hbox{$>$\kern-.75em\lower1ex\hbox{$\sim$}}}}
\def\la{\mathrel{\raise.3ex\hbox{$<$\kern-.75em\lower1ex\hbox{$\sim$}}}}

\def\I_M{{I_{\scriptscriptstyle M\times M}}}

\def\hypern{{_2F_{1}[\frac{n-2}{2(n-1)},\frac{1}{2},
     \frac{3n-4}{2(n-1)},
     -\frac{(n-1)(n-2)q^{2}}{2\beta^{2}r^{2n-2}}]}}

\def\hyperne{{_2F_{1}[\frac{n-2}{2(n-1)},\frac{1}{2},
     \frac{3n-4}{2(n-1)},
     -\frac{(n-1)(n-2)q^{2}}{2\beta^{2}r_{ex}^{2n-2}}]}}
\def\hypernp{{_2F_{1}[\frac{n-2}{2(n-1)},\frac{1}{2},
     \frac{3n-4}{2(n-1)},
     -\frac{(n-1)(n-2)q^{2}}{2\beta^{2}r_{+}^{2n-2}}]}}
\def\hypernx{{_2F_{1}[\frac{n-2}{2(n-1)},\frac{1}{2},
     \frac{3n-4}{2(n-1)},
     -\frac{(n-1)(n-2)x^{2}}{2\beta^{2}}]}}

\begin{document}
\begin{titlepage}
\begin{center}

\thispagestyle{empty}

\vskip 2cm \centerline{ \Large \bf A Note on Charged Black Holes in AdS space   
 }
\vskip .2cm \centerline{ \Large \bf and the Dual Gauge Theories}

\vskip .5in

{\bf Souvik Banerjee\footnote{e-mail: souvik@iopb.res.in}\\
\vskip .1in
{\em Institute of Physics,\\
Bhubaneswar 751005, India.}}

\vskip .2in
\end{center}

\begin{abstract}
We study the thermodynamics and the phase structures of Reissner-Nordstr\"{o}m and Born-Infeld black holes in AdS space by constructing ``off-shell'' free energies using thermodynamic quantities derived directly from the action. We then use these results to propose ``off-shell'' effective potentials for the respective boundary gauge theories. The saddle points of the potentials describe all the equilibrium phases of the gauge theories.
\end{abstract}
\vskip 1.2cm


\vskip 2in

\end{titlepage}
\vfill
\eject

\newpage
\setcounter{footnote}{0}
\section{INTRODUCTION}

\noindent It is by now evident that there exists a correspondence which relates 
gravity in Anti de Sitter space-time with a 
particular class of quantum field theories in one less 
dimension \cite{review, gkt}. This correspondence, related by duality,
has recently motivated  many researchers. This is due to the fact that 
one can begin addressing issues in quantum theory of
gravity via computations in weekly coupled field theories
and vice versa. A classic and well studied example of this 
type is the supergravity on $AdS_5 \times S^5$ which is dual to the ${\cal N} = 
4$ super Yang-Mills in four dimensions.
In general, for a $(n + 1 +q)$ dimensional theory of gravity compactified on 
$AdS_{n+1}\times X^q$, the dual field theory lives on a space whose topology is 
same as that of the boundary of $AdS_{n+1}$. The isometries of $X^q$ becomes 
global symmetries of the field theory. For example, when $X^q$ is a five-sphere, 
the $SO(6)$ isometry allows one to introduce three independent R-charges 
corresponding to the three Cartans of $SO(6)$. Consequently, one can turn
on three independent chemical potentials in ${\cal N} =4$ SYM. At finite 
temperature, the gravity dual of this theory is the R-charged black holes of 
${\cal N}=2$ gauged supergravity \cite{son1, cvetic, duff}. For the special case, when the charges are equal, 
these black holes reduce to the Reissner-Nordstr\"{o}m black holes in AdS space. Many features of
these black holes and their gauge theory duals were studied in \cite{emparan1,emparan2}.

\noindent Working at the supergravity level corresponds to analyzing gauge theories, at 
infinite coupling, with large number of colours. To see any finite 
coupling/finite colour effect in gauge theory, one requires studying string 
theory  on AdS. However, since this is as yet a poorly understood area, many 
authors have looked into the effects of adding $\alpha^\prime$ corrections to 
supergravity. See \cite{cvetic1,nojiri, brigante, buchel} for an incomplete list of references. In general, it is 
also expected that string theory will introduce higher order gauge field 
corrections to supergravity actions. These corrections, in turn, would modify 
various equilibrium and non-equilibrium properties of the gauge theory. 
See \cite{anindya} for work in this direction. At finite temperature, the gravity duals of
these are the black holes in the presence of higher derivative corrections. 
Construction of such black holes becomes progressively  difficult as one introduces 
more and more higher derivative terms in the action. In fact,  in many cases, 
one relies on perturbative construction of the black holes. However, there 
exists a  
rare example of exact black hole solution which takes into account a 
specific set of gauge field higher derivative corrections to all orders. These
are the black holes in the Born-Infeld (BI) theories in the presence of a negative 
cosmological constant. BI black holes were constructed in \cite{krug, tanay}. 
{\it Assuming} that there 
exists a dual gauge theory, equilibrium and non-equilibrium properties of the 
finite temperature gauge theory were studied by many authors by exploiting the 
black hole  solution\footnote{We have discussed previously that adding a 
gauge field in the bulk is equivalent
to turning of a chemical potential in the gauge theory. Since BI black holes 
accommodate all order gauge field corrections, they incorporate large 
chemical potential contributions into the gauge theory.}\cite{caisun, tan}. The purpose of the 
present work is to address some 
issues along these directions. The main aim of our work is to propose an off-shell effective potential for boundary gauge theory on $S^3$ at finite temperature and finite chemical potentials which on-shell reproduces various phases expected from $AdS$-CFT correspondence. These gauge theories are dual to Reissner-Nordstr\"{o}m and Born-Infeld black holes. Due to the non-availability of a systematic approach to work with strongly coupled theories, we take an indirect route. First, we construct an off-shell ``free-energy function'' in the bulk and then use it, along with $AdS$-CFT rules, to propose effective potentials for the dual gauge theories.

\noindent This paper is structured as follows. In the next section we review the Born-Infeld black hole solutions in $AdS$ space in $(n+1)$-dimensions. In section 3 we compute the Born-Infeld actions in two different thermodynamical ensembles - namely the fixed potential and the fixed charge ensembles. In the following section we compute different thermodynamic quantities in the two ensembles directly from the action. In section 5, we go into the study of phase structure of those black holes in grand canonical (fixed potential) ensemble. Although those have already been well studied \cite{krug, Fernando, cai}, we use a different mean field theory technique to find an off-shell potential known as the Bragg-Williams potential in condensed matter literature \cite{chaikin}. We start with an easier system, namely the Reissner-Nordstr\"{o}m, which is the zeroth order expansion of Born-Infeld solution and study its phase structure using Bragg-Williams construction. This shows a first order phase transition corresponding to the Hawking-Page phase transition from black hole phase to $AdS$ phase. We then repeat the same exercise for Born-Infeld case. After constructing the off-shell potentials in the gravity side, in section 6, we construct off-shell effective potentials for the boundary theory for both Reissner-Nordstr\"{o}m and Born-Infeld cases using $AdS$-CFT dictionary. Finally we study the phase structures thereof using the constructed effective potentials and again get first order phase transitions, which from the perspective of the boundary gauge theory, would now correspond to the confinement-deconfinement transitions. The appendix to this work includes a discussion on how using proper scaling one can go from a black hole geometry with elliptical horizon to one with flat horizon geometry and obtain thermodynamical quantities therein. This scaling, when applied on constructed Bragg-Williams potential, leads to the effective potentials for gauge theories on $R^3$. The example we have considered there is that of Reissner-Nordstr\"{o}m, for simplicity and we expect the same argument to hold for the Born-Infeld case as well.
\section{{\hspace{-.5cm} BORN-INFELD BLACK HOLES IN AdS SPACE }}

\noindent We start by reviewing some essential features of Born-Infeld action and its black hole solution. Let us consider the $(n+1)$ dimensional Einstein-Born-Infeld action with a
negative cosmological constant  $\Lambda$ of the form
\begin{equation}
{ S} = \frac{1}{16\pi G}{\int {d^{n+1}x}\sqrt{-g}\Big[({\cal R}-2\Lambda)+ 
L(F)\Big]}, 
\label{1eq1}
\end{equation}
where $ L(F)$  is given by
\begin{equation}
{ L(F)} = {4{\beta}^2\Big (1-\sqrt{1+\frac{F^{\mu \nu} 
F_{\mu \nu}}{2{\beta}^2}}\Big)}.
\label{}
\end{equation}
The constant $\beta$  is called the Born-Infeld parameter and has
the dimension of mass. In the limit $\beta \rightarrow \infty $ , higher order gauge field fluctuations can be neglected and, therefore, $ L(F)$ 
reduces to the standard Maxwell form
\begin{equation}
{ L(F)} = {- F^{\mu \nu} F_{\mu \nu}} +{\cal{O}}(F^4).
\label{}
\end{equation}
Thus the action, S, reduces to the standard form for which the Reissner-Nordstr\"{o}m in AdS is the black hole solution. Thermodynamics and phase structure of such black holes were studied in detail in \cite{emparan1,emparan2}. 
\footnote{In what follows, for simplicity, we will work in a unit in which $16 \pi G =1$, $G$ being the Newton's 
constant in $(n+1)$ dimensions.}

\noindent Equations of motions can be obtained by varying the action with respect to the gauge field $ A_\mu
$ and the metric  $ g_{\mu \nu} $. For $ A_\mu$ and for $ g_{\mu \nu} $ those are respectively given by

\begin{equation}
{{ \bigtriangledown}_\mu \Big(\frac{ F^{\mu \nu}}
{\sqrt{1+\frac{F^2}{2{\beta}^2}}}\Big)} ={0},
\label{2eq4}
\end{equation}
and 
\begin{equation}
{\cal R}_{\mu\nu}-\frac{1}{2}  {\cal R} g_{\mu\nu}+\Lambda
g_{\mu\nu}=\frac{1}{2}g_{\mu\nu}L(F)+ \frac{2F_{\mu
\alpha}F^{~\alpha}_{\nu}}{\sqrt{1+\frac{F^{\mu\nu}F_{\mu\nu}}{2\beta^{2}}}},
\label{2eq5}
\end{equation}
where ${\cal R}_{\mu\nu}$ is the Ricci tensor and ${\cal R}$, the Ricci scalar.
In order to solve the equations of motion, we use the metric 
ansatz
\begin{equation}
{ds^2} = {-V(r)dt^2 + \frac{dr^2}{V(r)} + f^{2}(r)g_{ij}dx^{i}dx^{j}},
\label{2eq6}
\end{equation}
The metric on the foliating submanifold, $g_{ij}$, is a function of coordinates $x^{i}$ and spans an $(n-1)$-dimensional hypersurface with scalar curvature  $(n-1)(n-2)k$, $k$ being a constant which characterizes the afore-mentioned hypersurface. Depending on whether the black hole horizon is elliptical, flat or hyperbolic, k can be taken as $\pm 1$ and $0$ respectively without any loss of generality. For the metric (\ref{2eq6}), we have non-vanishing
components of Ricci tensor
\begin{eqnarray}
\label{2eq7}
 && { \cal R}^{t}_{t}= -\frac{V''}{2} - (n-1)\frac{V'R'}{2R},
   \\
   \label{2eq8}
&& { \cal R}^{r}_{r}=
-\frac{V''}{2}-(n-1)\frac{V'R'}{2R}-(n-1)\frac{VR''}{R},
\\
\label{2eq9}
 && {\cal R}^{i}_{j}= \left( \frac{n-2}{R^2}k
-\frac{1}{(n-1)R^{n-1}}[V(R^{n-1})']'\right)\delta^{i}_{j},
\end{eqnarray}
where the primed quantities denote the derivatives with respect to $r$.

\noindent Let us consider the case where all the components of $F^{\mu\nu}$ are zero except $F^{rt}$. In that case (\ref{2eq4}) can be immediately solved to yield 
\begin{equation}
{F^{rt}} = {\frac{\sqrt{(n-1)(n-2)}\beta q}{\sqrt{2{\beta}^2
        {r}^{2n-2}+(n-1)(n-2)q^2}}}.
\label{emeq}
\end{equation}
Here $ q $ is an integration constant and is related to the electromagnetic charge. 
From (\ref{emeq}) we can also find the electric gauge potential as
\be
A_{t}={\frac{1}{c}}{ \frac{q}{r^{n-2}}~{}_2F_1\Big[ {{n-2}\over{2n-2}}, 
{1\over 2}, {3n-4\over{2n-2}}, 
-{\frac{(n-1)(n-2)q^2}{2{\beta}^2 r^{2n-2}}}\Big]}-\phi,
\label{gauge}
\ee
where $\phi$ is a constant and can be interpreted as the electrostatic potential difference between the black hole horizon and infinity and $c$ is a constant given by $c=\sqrt{\frac{2(n-2)}{n-1}}$. We choose $\phi$ in a way that makes $A_{t}$ vanish at the horizon\footnotetext[1]{Actually $A_{t}$ at the horizon $r=\rp$ cannot be chosen arbitrarily. The event horizon of the afore-mentioned background is a killing horizon of killing vector $\partial_t{}$ and therefore contains a bifurcation surface at $r=\rp$ where the killing vector vanishes. This in turn demands the vanishing of $A_{t}$ at $r=\rp$  if the one form $A$ is to be well-defined \cite{hartnoll,kobayashi}. A more detailed discussion regarding this can be found in \cite{wald1}.}. 
\bea
\phi=\frac{1}{c}\frac{q}{\rp^{n-2}}\hypernp.
\label{bipot}
\eea

\noindent Now if  $F^{rt}$ is the only non-zero component of all the $F^{\mu\nu}$'s, one can easily check from equation (5) that 
${\cal R}^{r}_{r}={\cal R}^{t}_{t}$ and hence, from (7) and (8) it follows $R''(r)=0$ which has two solutions, $f(r)=r$ and $f(r)=Constant$.  We will consider the case of $f(r)=r$ in this work. With this, and setting $\Lambda =-n(n-1)/2l^2$, we get the solution for $V(r)$ as \cite{tanay,cai}
\begin{eqnarray}
\label{12}
 V(r) &=&  k -\frac{m}{r^{n-2}} +\left(
\frac{4\beta^2}{n(n-1)} +\frac{1}{l^2}\right) r^2
 \nonumber \\
&& -\frac{2\sqrt{2} \beta}{(n-1)r^{n-2}}\int
\sqrt{2\beta^2r^{2n-2} +(n-1)(n-2)q^2 }dr.
\label{solmet}
\end{eqnarray}
$m$ here is an integration constant. Later we will see that this is related to the ADM mass of the black hole. The integral can also be expressed in terms of hypergeometric functions:
\begin{eqnarray}
\label{13}
V(r) &=&k-\frac{m}{r^{n-2}}+\left(\frac{4\beta^{2}}{n(n-1)}+\frac{1}{l^{2}}\right) r^{2}
    -\frac{2\sqrt{2}\beta\sqrt{2\beta^{2}r^{2n-2}+(n-1)(n-2)q^{2}}}{n(n-1)r^{n-3}}
      \nonumber \\
    &&+\frac{2(n-1)q^{2}}{nr^{2n-4} } \ _2F_{1}[\frac{n-2}{2(n-1)},\frac{1}{2},
    \frac{3n-4}{2(n-1)},
    -\frac{(n-1)(n-2)q^{2}}{2\beta^{2}r^{2n-2}}].
\label{v}
\end{eqnarray}
It is worth mentioning here that there is an ambiguity in the lower limit of the integral in the RHS of eqn.(\ref{solmet}). In order to fix this up, one has to invoke again the fact that $V(r)$ should reduce to that of Reissner-Nordstr\"{o}m \cite{emparan1} once $\beta \rightarrow \infty $ limit is taken. This tells that the lower limit of the integral should be such that the integral vanishes at that limit.\\

\noindent Black hole horizon satisfies $V(r)=0$. Denoting the solution as $r=r_+$, one can express $m$ in terms of $r_+$ as
\bea
m&=&r_+^{n-2}  +\Big[ {4 {\beta}^2\over{n(n-1)}}+ \frac{1}{l^2}\Big]r_+^n 
-{\frac {2 {\sqrt{2}}{\beta}r_+}{n(n-1)}}{\sqrt {2 \beta^2 r_+^{2n-2}+{(n-1)(n-2)q^2}}}\nonumber \\
&+&{\frac{2(n-1)q^2}{nr_+^{n-2}}}{}_2F_1\Big[ {{n-2}\over{2n-2}}, {1\over 2}, {{3n-4}\over{2n-2}}, 
-{\frac{(n-1)(n-2)q^2}{2 \beta^2 r_+^{2n-2}}}\Big].
\label{m}
\eea

\noindent Next, to find the temperature of the black hole, we follow the standard prescription and expand $V(r)$ in Taylor expansion around $r = r_+$ so that 

\begin{eqnarray}\nonumber
 V(r)\simeq\frac{\partial{V}}{\partial{r}}|_{r=r_+}(r-r_+) 
\end{eqnarray}
Using this and a redefinition of the variable ${r}$, the radial and temporal part of the metric reduces to the form 
\begin{equation}
{ds^2} = {d\rho^2 +\rho^2 d\left(\frac{\partial{V}}{\partial{r}}\Big\vert_{r=r_+}\frac{\tau}{2}\right)^2} 
\end{equation}
$\tau$ being the Euclidean time.
Now, to avoid conical singularity $\left(\frac{\partial{V}}{\partial{r}}\Big\vert_{r=r_+}\frac{\tau}{2}\right)$ should have a periodicity of $2\pi$ and the periodicity in $\tau$ is therefore given by 
\begin{eqnarray}\nonumber
\beta_{bh}=\frac{4\pi}{\frac{\partial{V}}{\partial{r}}\Big\vert_{r=r_+}}
\end{eqnarray}
This period is identified with the inverse of black hole temperature, $T_{bh}=\frac{1}{\beta_{bh}}$.\\\\
For our case ${\frac{\partial{V}}{\partial{r}}\Big\vert_{r=r_+}}$ can be easily found from eqn. (\ref{12}). Once again, one has to fix the lower limit of the integral and regarding this, the discussion at the end of eqn. (\ref{13}) still holds.
Finally the temperature of the black hole comes out to be
\be
T_{bh}= {{\frac{1}{4\pi}}\Big[{\frac{n-2}{r_+}}k +
 \Big\lbrace{\frac{4{\beta}^2}{n-1}}+\frac{n}{l^2} \Big 
  \rbrace{ r_+}-
{\frac{2{\sqrt2}\beta}{(n-1){r_+}^{n-2}}}{\sqrt{2{\beta}^2 
        {r_+}^{2n-2} + (n-1)(n-2)q^2}} \,\Big]},
\label{temp}
\ee
which matches exactly with the expression of temperature obtained in\cite{tanay, cai}.
From now on we will take $k=1$ for all our computations.

\noindent There are normally two ways to calculate thermodynamic quantities. First, one assumes that the black hole satisfies laws of thermodynamics and uses that to find thermodynamic quantities. Second is to compute the action for a black hole and use it to derive various state variables following standard prescription. Here we will follow the second path.

\section{{\hspace{-.5cm} ACTION CALCULATION }}
We will now calculate the black hole action in two different ensembles. Firstly we will focus on the grand canonical ensemble which is defined as a fixed potential ensemble. In the language of thermodynamics, this can be thought of as connecting the system to a heat reservoir full of quanta at a temperature, $T_{bh}$, the reservoir being identified as a pure AdS background with charged and uncharged quanta which are free to fluctuate in presence of a constant potential $\phi$. The scenario is quite different in case of a fixed charge, namely the canonical ensemble. Since AdS with localized charge is not a solution of BIAdS equation, pure AdS background cannot serve the purpose of a heat reservoir. It turns out that extremal black hole background is a good candidate in this regard.\footnote{This follows from an argument of \cite{emparan1} where the extremal black hole solution was used as a background on which the free energy was computed for canonical ensemble. We expect this to hold good for our finite $\beta$ case as well. } In order to keep charge, $Q$ fixed, we, in this case, retain only neutral quanta in the heat reservoir. \footnote{In grand canonical ensemble, an action calculation in four dimensions was performed earlier in \cite{Fernando}. We generalize the computation for arbitrary dimensions.}
\subsection{Fixed Potential}
The action for this is the one given in (\ref{1eq1}) analytically continued to Euclidean space by taking $t\rightarrow i\tau$. 
We then use the equation of motion given in (\ref{2eq5}) for the metric to eliminate $\cal R$ to obtain the on-shell action as:
\be
{ S} = {\int {d^{n+1}x}\sqrt{-g}\Big[\frac{4\Lambda}{n-1} - \frac{2L(F)}{n-1}- \frac{4F^2}{(n-1)}\frac{1}{\sqrt{1+\frac{F^2}{2\beta^2}}}
\Big]},
\label{onshell}
\ee
It is worth mentioning in this regard that since the space is asymptotically AdS, there is no contribution from the Gibbons-Hawking-York boundary term. Also the surface term that arises from the variation of the action with respect to the gauge field vanishes in this case, since, for this particular ensemble, the potential is kept fixed at $\infty$. Furthermore, since we contemplate on purely electrical solutions only (only non-zero component of $F^{\mu\nu}$ being $F^{r\tau}$), the possibility of having a Chern-Simons term does not arise as well.
\newline
\newline
Now we use the equation of motion for the gauge field given in (\ref{2eq4}) and get the full on-shell action as 
\be
I_{bh}=\omega_{n-1}\int^{\beta_{bh}}_0d\tau\int^\infty_{r_+} dr\Big[\frac{2n}{l^2}r^{n-1}+\frac{8\beta^2}{n-1}r^{n-1}-\frac{8\beta}{n-1}\sqrt{\beta^2r^{2n-2}+q^2\frac{(n-1)(n-2)}{2}}\Big],
\label{bh}
\ee 
$\omega_{n-1}$ being the volume of a unit $(n-1)$ sphere.
This integral is clearly divergent. This is because of the infinite volume of the black hole spacetime. This is where the idea of introducing a heat reservoir in form of background pure AdS spacetime, as discussed in the beginning of this section exactly fits in. What we would do is to subtract from (\ref{bh}) the pure AdS action,
\be
I_{AdS}=\omega_{n-1}\int^{\beta_{AdS}}_0d\tau\int^\infty_0 dr\Big[\frac{2n}{l^2}r^{n-1}\Big],
\ee 
which is also evidently infinity.\newline
In order to implement this regularization scheme \cite{witten} properly, we put an upper cut-off $R$ on the radial integration, which we would eventually take to infinity. For the black hole space-time to be smooth, $\beta_{bh}$ is given by the inverse of Hawking temperature, $T_{bh}$, given in eqn.(\ref{temp}). $\beta_{AdS}$ can, in general, be anything. But there is one constraint. $\beta_{AdS}$ should have the value which makes the geometries of the AdS and the black hole spacetimes the same on the asymptotic hypersurface defined by $r=R$. This is done by setting
\bea
\beta_{AdS}\sqrt{\Big[1+\frac{R^2}{l^2}\Big]}&=&\beta_{bh}\Big[1-\frac{m}{R^{n-2}}+\frac{4\beta^2}{n(n-1)}R^2+\frac{R^2}{l^2}\nonumber\\&-&\frac{2\sqrt{2}\beta}{n(n-1)R^{n-3}}\sqrt{2\beta^2R^{2n-2}+(n-1)(n-2)q^2}\nonumber\\
&+&\frac{2(n-1)q^2}{nR^{2n-4}}\hypern\Big]^{\frac{1}{2}}.
\eea
After some algebraic manipulation, this becomes, 
\bea
\beta_{AdS}&=&\beta_{bh}\Big[1-\frac{ml^2}{2R^n}+\frac{2\beta^2l^2}{n(n-1)}\{1-\sqrt{1+\frac{(n-1)(n-2)q^2}{2\beta^2R^{2n-2}}}\}\nonumber\\&+&\frac{(n-1)q^2l^2}{nR^{2n-2}}\hypern\Big].
\eea
Using this relation along with eqn..(\ref{m}) and then taking the limit $R\rightarrow\infty$ we finally get the Born-Infeld action in the grand canonical ensemble as
\bea
I_{GC}&=&\omega_{n-1}\beta_{bh}\Big[r_+^{n-2}-\frac{r_+^n}{l^2}-\frac{4\beta^2r_+^n}{n(n-1)}\{1-\sqrt{1+\frac{(n-1)(n-2)q^2}{2\beta^2r_+^{2n-2}}}\}\nonumber\\&-&\frac{2(n-1)}{n}q^2\frac{1}{r_+^{n-2}}\hypern\Big].
\label{gc}
\eea
\\
As a consistency check of our result, we see that with  $\beta\rightarrow\infty$ limit,
\bea
I_{GC}\Big\vert_{\beta\to \infty}=\omega_{n-1}\beta_{bh}\Big[\rp^{n-2}-\frac{\rp^n}{l^2}-\frac{q^2}{\rp^{n-2}}\Big].
\label{rngc}
\eea
which is exactly the same as the Reissner-Nordstr\"{o}m action for the grand canonical ensemble as obtained in \cite{emparan1}.

\subsection{Fixed Charge}
In this ensemble, we, instead of fixing the potential at infinity, fix the charge of the black hole. Then the action given in (\ref{onshell}) is no longer the appropriate one. Since the potential is not fixed at infinity, the boundary term as obtained by the variation of the gauge field, unlike in the case of fixed potential ensemble, has a non-vanishing contribution given by 
\be
I_{s}= -4\int d^{n}x \sqrt{-h}\frac{F_{\mu\nu}}{\sqrt{1+\frac{F^2}{2\beta^2}}}n_{\mu}A_{\nu},
\ee
which after some straightforward computation becomes
\be
I_{s}= 2(n-1)\omega_{n-1}\beta_{bh}\frac{q}{{r_+}^{n-2}} \hypern,
\ee
$h$ being the determinant of the induced metric at the boundary and $n_{\mu}$, the radial unit vector pointing outward. Not only that, we also have to subtract the pure AdS background as before to ensure the convergence of the integral, the difference with the previous case of fixed potential ensemble being only that in the present case AdS background cannot be interpreted as the metric background or heat reservoir as argued in the beginning of the section.
\bea
I_{bh}+I_{s}-I_{AdS}&=& \omega_{n-1}\beta_{bh}\Big[\rp^{n-2}-\{\frac{\rp^n}{l^2}+4\beta^2\rp^n\}+\frac{2\sqrt{2}\beta\rp}{n(n-1)}\sqrt{2\beta^2\rp^{2n-2}+(n-1)(n-2)q^2}\nonumber\\&+&\frac{2(n-1)^2q^2}{n\rp^{n-2}}\hypern\Big].
\label{can1}
\eea
The metric background in this case is the extremal black hole. The action for the extremal black hole can be found by substituting in (\ref{can1}), the condition for extremality with $\rp=\re$, $\re$ being the horizon of the extremal black hole.\newline
The condition for extremality can be obtained by setting $T_{bh}=0$ as
\bea
(n-2)\re^{n-3}+\Big[\frac{n}{l^2}+\frac{4\beta^2}{n-1}\Big]\re^{n-1}-\frac{2\sqrt{2}\beta}{n-1}\sqrt{2\beta^2\re^{2n-2}+(n-1)(n-2)q^2}=0.
\label{condex}
\eea
And with this the action for the extremal black hole becomes
\be
I_{ex}=2(n-1)\omega_{n-1}\beta_{bh}\Big[\frac{\re^{n-2}}{n}+\frac{(n-1)q^2}{n\re^{n-2}}\hyperne\Big].
\ee
Subtracting the extremal background, finally, the full Born-Infeld action for canonical ensemble becomes:
\bea
I_{C}&=&\omega_{n-1}\beta_{bh}\Big[\rp^{n-2}-\{\frac{\rp^n}{l^2}+\frac{4\beta^2\rp^n}{n(n-1)}\}+\frac{2\sqrt{2}\beta\rp}{n(n-1)}\sqrt{2\beta^2\rp^{2n-2}+(n-1)(n-2)q^2}\nonumber\\&+&\frac{2(n-1)^2q^2}{n\rp^{n-2}}\hypern\nonumber\\&-&2(n-1)\omega_{n-1}\beta_{bh}\{\frac{\re^{n-2}}{n}\nonumber\\&+&\frac{(n-1)q^2}{n\re^{n-2}}\hyperne\}\Big].
\label{can}
\eea
\\

As a check of our computation, if we take $\beta\rightarrow\infty$ limit of (\ref{can}) we get,
\bea
I_{C}\Big\vert_{\beta\to \infty}=\omega_{n-1}\beta_{bh}\Big[\rp^{n-2}-\frac{\rp^n}{l^2}-\frac{(2n-3)q^2}{\rp^{n-2}}-\frac{2(n-1)}{n}\re-\frac{2(n-1)^2}{n}\frac{q^2}{\re^{n-2}}\Big],
\label{rnc}
\eea
which is exactly the same as the Reissner-Nordstr\"{o}m action as obtained for the fixed charge ensemble in \cite{emparan1}.
In the next section, we calculate thermodynamic quantities directly from those actions. 

\section{{\hspace{-.5cm} THERMODYNAMICAL QUANTITIES }}
The state variables for the system can be computed from the actions, $I_{GC}$ and $I_{C}$ given in (\ref{gc}) and (\ref{can}) respectively.

\subsection{Fixed Potential}
The grand canonical free energy is given by $F_{GC}=E-TS-Q\phi$. Now F is also equal to $\frac{I_{GC}}{\beta_{bh}}$. Combining these two definitions we can find the state variables for the system as follows: 
\bea
E&=&\Big(\frac{\partial{I_{GC}}}{\partial{\beta_{bh}}}\Big)_{\phi}-\frac{\phi}{\beta_{bh}}\Big(\frac{\partial{I_{GC}}}{\partial{\beta_{bh}}}\Big)_{\beta},\\
S&=&\beta_{bh}\Big(\frac{\partial{I_{GC}}}{\partial{\beta_{bh}}}\Big)_{\phi}-I_{GC},\\
Q&=&-\frac{1}{\beta_{bh}}\Big(\frac{\partial{I_{GC}}}{\partial{\beta_{bh}}}\Big)_{\phi}.
\label{thermo}
\eea
Now for this ensemble, $\phi$ is a constant. Thus to find the partial derivatives keeping $\phi$ constant, one has to substitute the condition $\frac{\partial{\phi}}{\partial{\rp}}=0$, which we obtain from (\ref{gauge}) keeping in mind that in this case $q$ is no longer a constant, but a function of $r_+$.\\
With all these, we get the state variables as:
\bea
E&=&\omega_{n-1}(n-1)\Big[\rp^{n-2}+\{\frac{\rp^n}{l^2}+\frac{4\beta^2\rp^n}{n(n-1)}\}-\frac{2\sqrt{2}\beta\rp}{n(n-1)}\sqrt{2\beta^2\rp^{2n-2}+(n-1)(n-2)q^2}\nonumber\\&+&\frac{2(n-1)q^2}{n\rp^{n-2}}\hypernp\Big],
\label{egc}
\eea
using (\ref{m}), which can also write this as
\bea
E=\omega_{n-1}(n-1)m,
\eea
and
\bea
S&=&4\pi\omega_{n-1}\rp^{n-1},\label{thermo2}\\
Q&=&2\sqrt{2(n-1)(n-2)}\omega_{n-1}q.
\label{thermo1}
\eea
\subsection{Fixed Charge}
In the canonical ensemble, the free energy is given by $F_{C}=E-TS$, which is again equal to $\frac{I_{C}}{\beta_{bh}}$.
Then in a similar way as done before, one can find the corresponding state variables as:
\bea
E&=&\Big(\frac{\partial{I_{C}}}{\partial{\beta_{bh}}}\Big)_{q}=(n-1)m-(n-1)m_{ex},\\
S&=&\beta_{bh}\Big(\frac{\partial{I_{C}}}{\partial{\beta_{bh}}}\Big)_{q}-I_{C}=4\pi\omega_{n-1}\rp^{n-1},
\eea
where $m_{ex}$ is given by
\bea
m_{ex}&=&2\Big[\frac{\re^{n-2}}{n}+\frac{(n-1)q^2}{n\re^{n-2}}\hyperne\Big].
\eea
This expression for $m_{ex}$ can also be obtained by plugging in (\ref{m}) the condition for extremality, (\ref{condex}).

\noindent Having obtained the thermodynamical quantities, we would like to study various stable, unstable and metastable phases associated with the black hole. For that, we construct an ``off-shell'' free energy, the saddle points of which dictates the (in)stability of the black hole. The details of this construction is discussed in the next section. 
\section{{\hspace{-.5cm} CONSTRUCTION OF BRAGG-WILLIAMS FREE ENERGY \& STUDY OF PHASE STRUCTURE }}
A novel way to study the phase structure of a thermodynamical system is to construct a Landau free-energy. This free-energy is generally a function of order parameter and also depends on some intensive parameters like temperature, chemical potential etc. Various phases of the system appear via extremization of this free-energy function in terms of the order parameter. There is, however, another way to analyze stable, unstable or metastable phases of the system. This is known as the construction of the Bragg-Williams potential \cite{chaikin, sudipta}. This is particularly useful when transition from one phase to another phase involves a finite change in the order parameter. For our purpose, we find it suitable to use the Bragg-Williams construction. In the case of $\beta\rightarrow\infty$, i.e. for Reissner-Nordstr\"{o}m black hole, we know from \cite{emparan1} that there is a first order Hawking-Page (HP) transition. At a critical temperature, the black hole becomes unstable. The system prefers the AdS phase. This transition is of first order in nature, marked by a discontinuous change in the gravitational entropy. Our primary motivation would be to study the fate of this transition when $\beta$ is finite. So we would be interested in constructing Bragg-Williams potential for Born-Infeld black hole. In order to do so, we have to first decide on an order parameter.
To this end, we note that a first order phase transition is characterized by a discrete jump of the order parameter. In our case this jump shows up in the horizon radius of the black hole. Indeed, at the Hawking-Page (HP) point, AdS phase (identified with $\rp=0$) crosses over to the black hole phase (with non-zero $\rp$). So we find it suitable to use $\rp$ as the order parameter. Before we go on to discuss the phase structure in the Born-Infeld theory, we find it instructive to first analyze the Reissner-Nordstr\"{o}m case. In a later sub-section, we generalize this for Born-Infeld black holes. We, further, stick to the grand canonical ensemble for the rest of our discussions.
\subsection{Reissner-Nordstr\"{o}m }
The Bragg-Williams free energy for a Reissner-Nordstr\"{o}m black hole in a grand canonical ensemble is given by $W_{GC}=E-TS-Q\phi$ with $T$ and $\phi$ treated as external parameters. $E$ can be found by taking $\beta\rightarrow\infty$ limit of (\ref{egc}) with the understanding that since we are working in a fixed potential ensemble we have to write $q$ in terms of $\phi$. In order to achieve this we use the relation between charge and potential of Reissner-Nordstr\"{o}m black hole,
\bea
\phi=\frac{1}{c}\frac{q}{\rp^{n-2}},
\label{prn} 
\eea
which can be directly obtained by taking $\beta\rightarrow\infty$ limit of eqn.(\ref{bipot}).

\noindent With this, the Bragg-Williams free energy for the Reissner-Nordstr\"{o}m black hole is given by
\bea
W_{BW}^{RN}&=&E-TS-Q\phi \nonumber\\&=&\omega_{n-1}\Big[(n-1)\rp^{n-2}(1-c^2{\phi}^2)-4\pi\rp^{n-1}T+\frac{\rp^n}{l^2}(n-1)\Big].
\label{rnbw}
\eea
The on-shell temperature can be computed by differentiating $W_{BW}^{RN}$ with respect to $\rp$ and then setting it to zero. The temperature comes out to be \footnote{In eqn.(\ref{rnbw}), $\rp$ should be treated as an unconstrained variable. Only on shell, $\rp$ is related to $\phi$ and $T$. This can be found by inverting eqn.(\ref{temprn}) for $\rp$. }
\be
T_{RN}=\frac{(n-2) l^2 (1-c^2\phi^2)+n\rp^2}{4\pi l^2 \rp},
\label{temprn}
\ee
which is the same as the $\beta\rightarrow\infty$ limit of (\ref{temp}) and also matches with the expression for the temperature of Reissner-Nordstr\"{o}m black holes obtained in \cite{emparan1}.
The behaviour of $W_{BW}^{RN}$ as a function of the order parameter for a fixed $\phi$ and for different temperatures is shown in the figure. 
 
\begin{figure}[h]
\centering
\includegraphics[width=13cm,height=7cm]{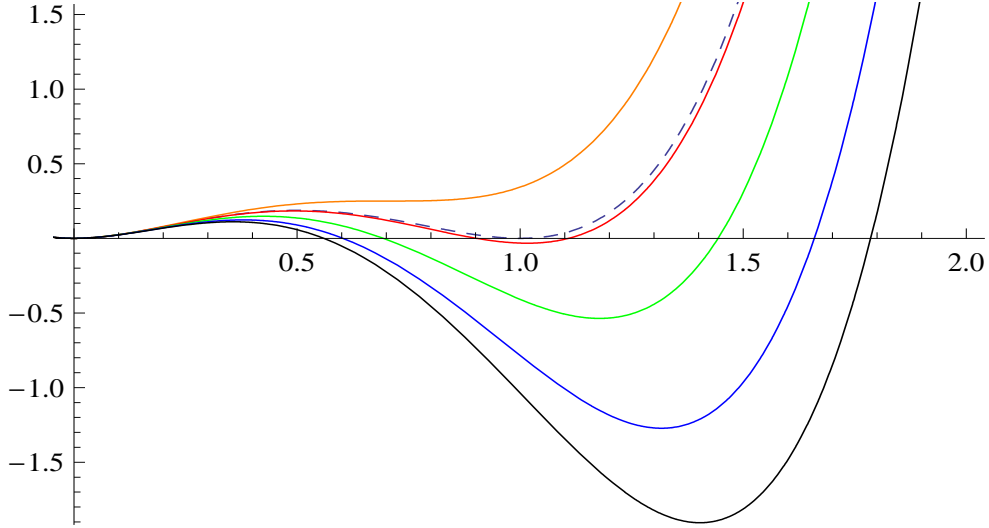}
\caption{This is a plot of $W_{BW}^{RN}$ as a function of $\rp$ for fixed $\phi$. The phase structure shown here is for $n=4$ and for $\phi$=0.0003. The dashed line is for the critical temperature, $T=T_c$, the orange one is the transition involving a metastable phase, another feature of a generic first order phase transition. The red, green, blue and black lines are for $T>T_c$ in an increasing order.}
\end{figure}

\noindent We see from the phase diagram that, the $\rp^3$ term present in the free energy expression for $n=4$ brings in an asymmetry in $W_{BW}^{RN}$ as a function of $\rp$ and results in an emergence of a secondary minimum at finite value of $\rp$. The value of $W_{BW}^{RN}$ at this secondary minimum is greater than zero when $T<T_c$, but becomes zero at the critical temperature $T=T_c$. For all $T>T_c$, $W_{BW}^{RN}$ is negative at the secondary minimum. Thus there is a phase transition from black hole to AdS as we tune the temperature below $T_c$, with a discontinuous change in ${\rp}$ at $T=T_c$.  This is, clearly, the signature of a first order phase transition occurring at $T=T_c$.\\
\\
An analytic expression for $T_c$ can be obtained on requiring that $W_{BW}^{RN}$ is an extremum with respect to $\rp$ in equilibrium, i.e, $\Big(\frac{\partial{W_{BW}^{RN}}}{\partial{\rp}}\Big)=0$ along with the condition that the free energies of the ordered and the disordered phases match exactly at the transition, which, in turn, implies, ${W_{BW}^{RN}}=0$. From these two conditions, we obtain the critical value of the order parameter, $\rp$. 

\noindent For Reissner-Nordstr\"{o}m case, in $(n+1)$ dimensions, the requirement, ${W_{BW}^{RN}}=0$ gives
\be
\rp = \frac{4\pi l^2 T+\sqrt{16l^4T^2\pi^2-4l^2(n-1)^2(1-c^2\phi^2)}}{2(n-1)}.
\label{zr}
\ee
The other one, namely $\Big(\frac{\partial{W_{BW}^{RN}}}{\partial{\rp}}\Big)=0$ gives
\be
\rp = \frac{4\pi l^2 T+\sqrt{16l^4T^2\pi^2-4l^2(n-2)(1-c^2\phi^2)}}{2n}.
\label{derzr}
\ee
Equations (\ref{zr}) and (\ref{derzr}) can be solved to yield the transition temperature, $T_c$ in terms of the corresponding critical value of $\phi$
\be
T_c=\frac{(n-1)}{2\pi l}\sqrt{1-c^2\phi_c^2}.
\label{tcfc}
\ee
This is precisely the same critical temperature, $T_c$ as obtained from the ${W_{BW}^{RN}}$ vs $\rp$ diagram, as expected.
\\
\\
\begin{center}
\begin{figure}[h]
\includegraphics[width=6.6cm]{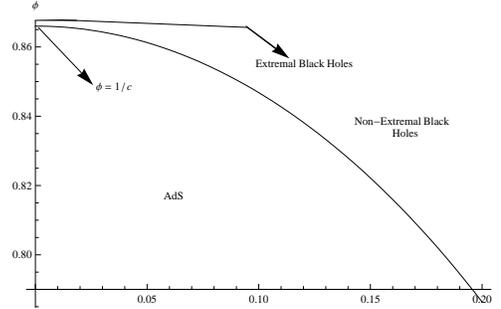}
\vspace{-1.3cm}
\caption{The phase structure of Reissner-Nordstr\"{o}m in fixed potential ensembles. $T=0$ line corresponds to extremal black holes. The extremal black holes are unstable. This plot is for $n=4$ and we have set $l=1$ here. }
\label{phiT} 
\end{figure}
\end{center}
A similar exercise can also be done keeping $T$ fixed and studying the phase structure varying the parameter, $\phi$. The resulting phase structure is shown in the diagram below. 

\begin{figure}[h]
\centering
\includegraphics[width=7cm]{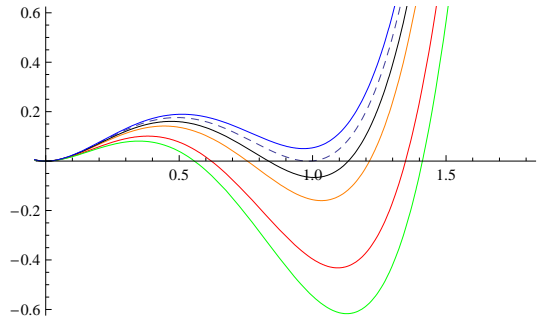}
\caption{This is a plot of $W_{BW}^{RN}$ as a function of $\rp$ for fixed $T$. The phase structure shown here is for $n=4$ and for $T=0.47$. The dashed line is for the critical value of potential, $\phi=\phi_c$, the blue one is for $\phi<\phi_c$. The black, orange, red and green lines are  for $\phi>\phi_c$ in an increasing order.}
\end{figure}
\noindent The behaviour shows, as expected, the features of first order phase transition at $\phi=\phi_c$. The analytic expression for $\phi=\phi_c$ can be obtained from eqn.(\ref{tcfc}) as 
\be
\phi_c = \frac{1}{c}\sqrt{1-\frac{4\pi^2 l^2 T_c^2}{(n-1)^2}}.
\label{fctc}
\ee 
The full phase structure in $\phi-T$ plane is shown in fig.\ref{phiT}.
Having discussed the $\beta\rightarrow\infty$ case, in the next sub-section we turn our attention to finite $\beta$.
 
\subsection{BORN-INFELD}
\begin{figure}[h]
\centering
\includegraphics[width=9cm]{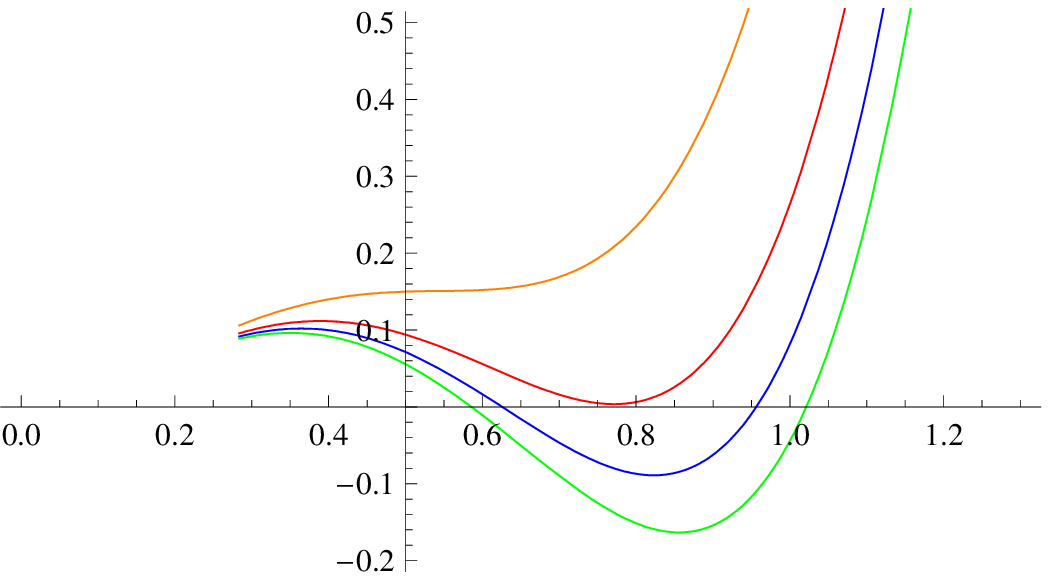}
\caption{$W_{BW}^{GC}$ is plotted against $\rp$ using $x$ as a parameter for $n=4$. We have fixed $\phi=0.2$ and have plotted for different values of temperature. The red line is for $T=T_c$. The blue and green lines are for $T>T_c$ in an increasing order, whereas the orange line is for $T<T_c$ }
\label{bi} 
\end{figure}

It turns out, owing to the non-linear relation between $\phi$ and $q$ as in eqn.(\ref{bipot}), a complete analytical treatment is difficult in this case. One way to circumvent this problem is to make large $\beta$ expansion and introduce $\frac{1}{\beta}$ correction order by order over the Reissner-Nordstr\"{o}m construction. However, this would not allow us to study the Bragg-Williams potential at finite $\beta$. So we restrict ourselves to a semi-analytic approach to construct the free energy. This is done as follows.
First we define a new variable, $x$ as\\
\bea
x=\frac{q}{{\rp}^{n-1}}.
\label{par}
\eea
The horizon radius, $\rp$ can now be rewritten as
\bea
\rp(x,\phi)=\frac{\phi{c}}{x\hypernx}.
\label{substx}
\eea
We can write down the grand canonical Bragg-Williams free energy for a Born-Infeld black hole as\\
\bea
W_{BW}^{GC}=E-TS-\phi{Q},
\label{frbw}
\eea
where $E$, $S$ and $Q$ are given by (\ref{egc}) with the substitution (\ref{substx}) being taken care of.
To see how $W_{BW}^{GC}$ behaves with change in the order parameter, we, therefore, do a parametric plot. The behaviour is shown in fig.\ref{bi}\footnote{Those plots go down smoothly to $\rp=0$ as in the case of Reissner-Nordstr\"{o}m. But unfortunately, that feature is not clearly visible in this phase diagram because of the fact that, the parameter, x, we have chosen for plotting goes as ~$\frac{1}{r}$. However, this feature can be easily checked from the expression for free energy directly. }, which again shows a first order phase transition at a critical temperature, $T_c$.
\\
We would like to mention one point in this regard. For the Reissner-Nordstr\"{o}m, in grand canonical ensemble, we would observe this phase structure only when $\phi{c}<1$ \cite{emparan1}. For Born-Infeld case also there is a similar critical value for $\phi{c}$ which can be determined by plotting the on-shell free energy against $T$ for different values of $\phi$\cite{Fernando}.

\section{{\hspace{-.5cm} PROPOSAL FOR EFFECTIVE POTENTIALS IN THE BOUNDARY THEORY }}
Assuming a gauge theory dual to the Born-Infeld black hole, in this section we will study some properties of this theory. In particular, we ask the following question: Can we atleast phenomenologically construct an effective potential in the gauge theory which describe its equilibrium properties? Since bulk has electrical charges, gauge theory in question must also have associated R-charges and corresponding chemical potentials.    

\noindent As we have discussed in the introduction, direct computation of effective potential in terms of the order parameter (the R-charge) in gauge theory is difficult. However, it is possible to use AdS/CFT conjecture and our computations in the previous sections to propose an effective potential whose saddle points represent various phases of the gauge theory. However, we should emphasize that the potential constructed this way may not be unique, except perhaps close to the transition line.\\\\
In the following, we first deal with the simpler case of gauge theory dual of Reissner-Nordstr\"{o}m black hole. Finally we generalize it to the Born-Infeld case.
\subsection{Reissner-Nordstr\"{o}m}
\begin{figure}[!]
\begin{minipage}[t]{8cm}
\vspace{-10pt}
\centerline{\hspace{6.3mm}
\rotatebox{0}{\epsfxsize=8cm\epsfbox{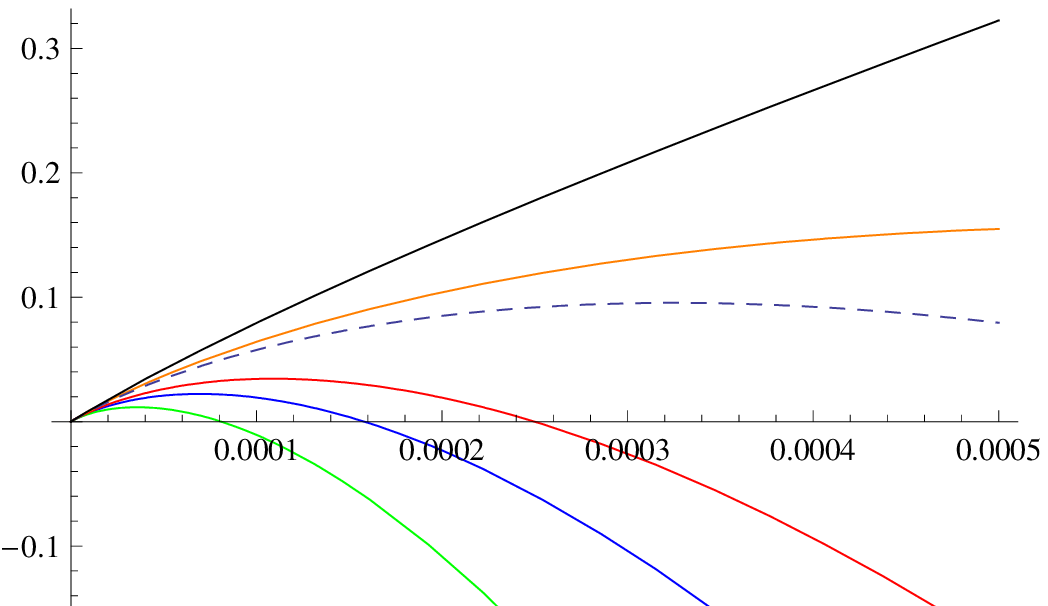}}}
\hspace{3.3cm}
\end{minipage}
\hfill
\begin{minipage}[t]{8cm}
\vspace{-10pt}
\centerline{\hspace{11.3mm}
\rotatebox{0}{\epsfxsize=8cm\epsfbox{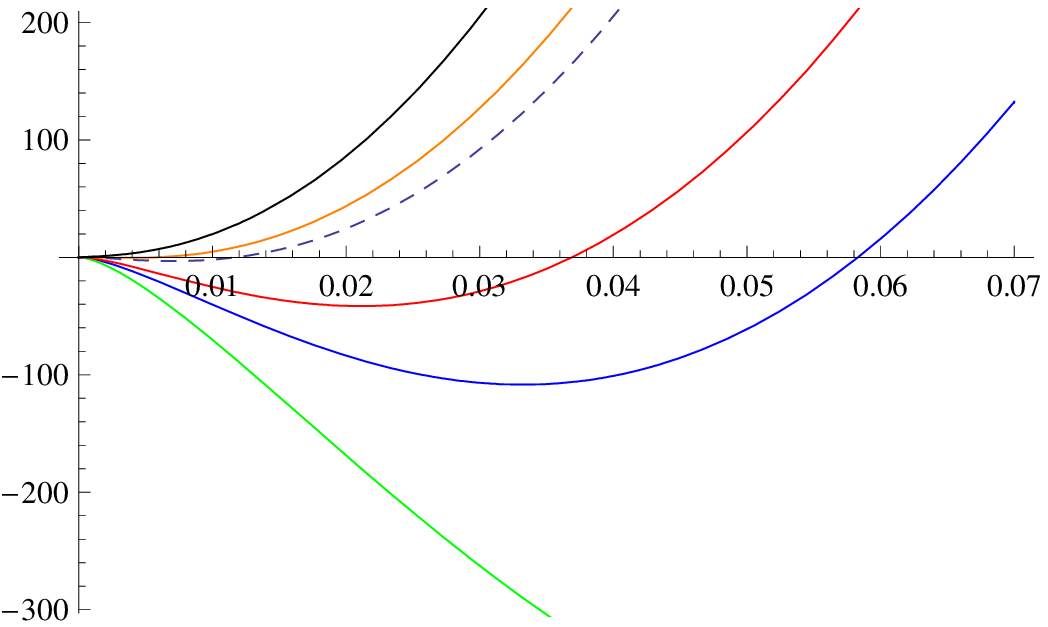}}}
\hspace{3.3cm}
\protect\label{fig3}
\end{minipage}
\caption[]{Plots of $W_{BT}^{RN}$ vs order parameter, $Q$ (the left one) for small Q values and (the right one) with relatively large range of values for Q for $n=4$, show the signature of first order phase transition. The dashed line is for $T=T_c$. The orange and the black lines are for $T<T_c$ in decreasing order in temperature, and the red, the blue and the green lines are for $T>T_c$ in increasing order in temperature. For both the plots $\phi$ is kept fixed at the value 0.03. We have also taken $N_c=1$, $\omega_3=1$ and $l=1$ while plotting these.}
\protect\label{fig4}
\end{figure}

While in the gravity theory the order parameter was $r_+$, in the dual theory the corresponding order parameter would be the physical charge, $Q=\int^*F$, which turns out to be the same as the charge one derives from the action. In our case, $Q=\omega_{n-1}\frac{1}{8\pi{G}}\sqrt{(n-1)(n-2)}q$, where $q$ is the ``charge'' that appears in the action and $\omega_{n-1}$, the $n-1$ dimensional transverse volume.\\
The conjugate chemical potential $\mu$ is the same as the electric potential, $\phi$ at the horizon given eqn.(\ref{prn}). In $n+1$ dimensions,
\be 
\mu=\phi=\sqrt{\frac{n-1}{2(n-2)}}\frac{q}{\rp^{n-2}}=\frac{4\pi{G}Q}{(n-2)\omega_{n-1}{\rp^{n-2}}}.
\label{rqprn}
\ee
We now use (\ref{rqprn}) to express $W_{BW}^{RN}$ given in (\ref{rnbw}) in terms of Q and $\phi$ in the following form\\
\bea
{W_{BT}^{RN}} &=& \frac{N_c^2}{8\pi^2}\omega_{n-1}\Big[\frac{2\pi^2 (n-1) (1-c^2\phi^2)}{(n-2)}\frac{Q}{\phi}- \frac{2^{\frac{3n-5}{n-2}}\pi^{\frac{3n-4}{n-2}}T}{(n-2)^{\frac{n-1}{n-2}}} \Big(\frac{Q}{\phi}\Big)^{\frac{n-1}{n-2}}\nonumber \\
   &+& \frac{2^{\frac{n}{n-2}}\pi^{\frac{2n}{n-2}}(n-1)}{(n-2)^{\frac{n}{n-2}}l^2}\Big(\frac{Q}{\phi}\Big)^{\frac{n}{n-2}}\Big],
\label{btfern}
\eea
where $Q$ is rescaled as $Q=\frac{Q}{{N_c}^2 \omega_{n-1}}$, $N_c$ being the number of colours. The motivation behind doing this scaling is that in the deconfined phase, the free energy and the charge, both are of the order of $N_c^2$. Therefore, the appropriate observable in large $N_c$ limit, is, instead of $Q$, $\lim_{N_c\to \infty}\frac{Q}{N_c^2}$.  We have also used the relation $G=\frac{{\pi}l^{n-1}}{2{N_c}^2}$ and while using this in the expression for effective potential, we have made it dimensionless by redefining $G$ as $\frac{G}{l^{n-1}}$.
The plot of the boundary effective potential ${W_{BT}^{RN}}$ given in (\ref{btfern}) against the new order parameter $Q$ again gives a first order phase transition as shown in fig.\ref{fig4}. This phase transition corresponds to the confinement-deconfinement transition in the strongly coupled gauge theory as discussed in \cite{witten}.

\noindent The temperature of the gauge theory can be found by extremizing $W_{BW}^{RN}$ with respect to the order parameter, $Q$ and this comes out to be 
\bea
T&=&\frac{ (n-2)^2 2^{-\frac{3n-5}{n-2}}\pi ^{-\frac{3n-4}{n-2}}}{n-1}
   \Big(\frac{Q}{(n-2)\phi}\Big)^{-\frac{1}{n-2}} \phi \Big[\frac{2(n-1)\pi^2}{(n-2) \phi }(1-c^2\phi^2)\nonumber \\
&+&\frac{n (n-1)}{l^2 (n-2)^2 \phi
   }2^{\frac{n}{n-2}} \pi ^{\frac{2n}{n-2}} 
   \left(\frac{Q}{(n-2) \phi }\right)^{\frac{2}{n-2}}\Big],
\label{soldw}
\eea
which is exactly the same as the Reissner-Nordstr\"{o}m temperature as in (\ref{temprn}) once we substitute in it $Q$ in terms of $\rp$ and $\phi$ through eqn.(\ref{rqprn}).\\
Following the discussion in section 6.1, we would now try to find the confinement-deconfinement transition temperature, $T_c$. The condition $W_{BT}^{RN}=0$ gives
 \bea
 T &=& 2 \pi  \Big[2^{\frac{1}{n-2}} \pi ^{\frac{2}{n-2}}
    \left(\frac{Q}{(n-2) \phi }\right)^{\frac{1}{n-2}}\Big]^{1-n}
    \Big[\frac{(n-1) \left(1-c^2 \phi ^2\right) 
    \left(2^{\frac{1}{n-2}} \pi ^{\frac{2}{n-2}} \left(\frac{Q}{(n-2) \phi
    }\right)^{\frac{1}{n-2}}\right)^{n-2}}{8 \pi ^2} \nonumber \\
 &+&\frac{(n-1) \left(2^{\frac{1}{n-2}} \pi ^{\frac{2}{n-2}}
    \left(\frac{Q}{(n-2) \phi }\right)^{\frac{1}{n-2}}\right)^n}{8 l^2 \pi
    ^2}\Big],
 \label{solw}
 \eea
whereas, the other requirement,  $\Big(\frac{\partial{W_{BT}^{RN}}}{\partial{Q}}\Big)=0$ gives (\ref{soldw}).
From (\ref{solw}) and (\ref{soldw}), we can find an equation involving critical charge, $Q_c$ as
\be
2^{\frac{2}{n-2}} \pi ^{\frac{4}{n-2}} \left(\frac{Q_c}{(n-2) \phi
   }\right)^{\frac{2}{n-2}}-l^2+c^2 l^2 \phi_c ^2=0.
\ee
Substituting this relation in (\ref{solw}) or (\ref{soldw}) we can write down the critical temperature, $T_c$ for the confinement-deconfinement transition as
\be
T_c=\frac{(n-1)}{2\pi l}\sqrt{1-c^2\phi_c^2},
\label{tcfc1}
\ee
which turns out to be exactly the same as that obtained in (\ref{tcfc}).

\subsection{Born-Infeld}
One can generalize the ideas mentioned in the previous sub-section to the case of Born-Infeld to find a gauge theory effective potential. But because of the non-linear non-invertible relationship between the electric potential at the horizon, $\phi$ and the charge, $Q$ as in eqn.(\ref{bipot}), it is not possible to write an exact expression for $\rp$ in terms of $Q$ and $\phi$. However, a parametric plot suggests that our construction leads us to a candidate effective potential for Born-Infeld dual.
Following the case of Reissner-Nordstr\"{o}m, we propose, in this case, the gauge theory effective potential, in $n=4$ as

\begin{figure}[h]
\centering
\includegraphics[width=6cm]{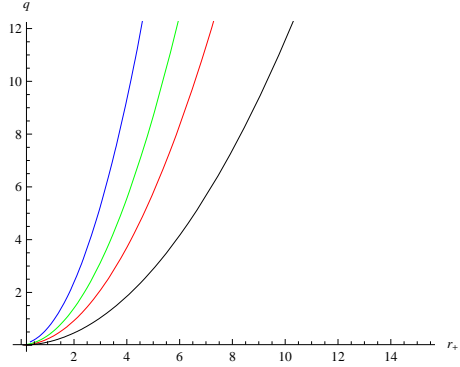}
\caption{Parametric plot of $q$ against $\rp$ for different values of the parameter, $\phi$. }
\label{qp} 
\end{figure}
\newpage
\bea
W_{BT}^{BI} &=& N_c^2 \omega_3\Big[ -\frac{T   \rp^3}{2 \pi }-  Q \phi  +\frac{3}{8 \pi ^2} 
   \Big(\frac{\beta^2 \rp^4}{3}+ \frac{\rp^4}{l^2}+ \rp^2- \frac{\beta \rp \sqrt{2 \beta^2 \rp^6 + 8
   \pi ^4 Q^2}}{3 \sqrt{2}} \nonumber\\
&+& \frac{2 \pi ^4 Q^2 \,
   _2F_1\Big(\frac{1}{3},\frac{1}{2},\frac{4}{3},-\frac{4 \pi ^4 Q^2}{\beta^2
   \rp^6}\Big)}{\rp^2}\Big)\Big],
\label{frbt}
\eea
along with the relation among chemical potential, $\mu$, charge, $q$ and $\rp$ from which one has to express $\rp$ in terms of $\mu$ and $q$. 
\be
\mu=\phi=\frac{\sqrt{3} q}{2 \rp^2}\ _2F_1\Big[\frac{1}{3}, \frac{1}{2}, \frac{4}{3}, -\frac{3 q^2}{\beta^2\rp^6}\Big],
\ee
where $q$ is the ``charge'' appearing at the action which can be related to the physical charge, ``$Q$'' through the relation given in eqn. (\ref{thermo1}). One can solve this equation numerically to find a relation between $\rp$ and $q$ for a fixed value of the parameter, $\phi$.

\noindent Equation (\ref{frbt}) is derived from (\ref{frbw}) by first substituting in it the expressions for $E$, $S$ and $Q$ given in equations (\ref{egc}), (\ref{thermo2}) and (\ref{thermo1}) with reinstatement of the gravitational constant, $G$ for $n=4$. We then use the relation $G=\frac{{\pi}l^3}{2{N_c}^2}$. However, we make $G$ dimensionless by dividing it by $l^3$ and scale Q as $\frac{Q}{N_c^2}$ for the same reason as given in the previous section in the context of Reissner-Nordstr\"{o}m.
\begin{figure}[h]
\centering
\includegraphics[width=10cm,height=10cm]{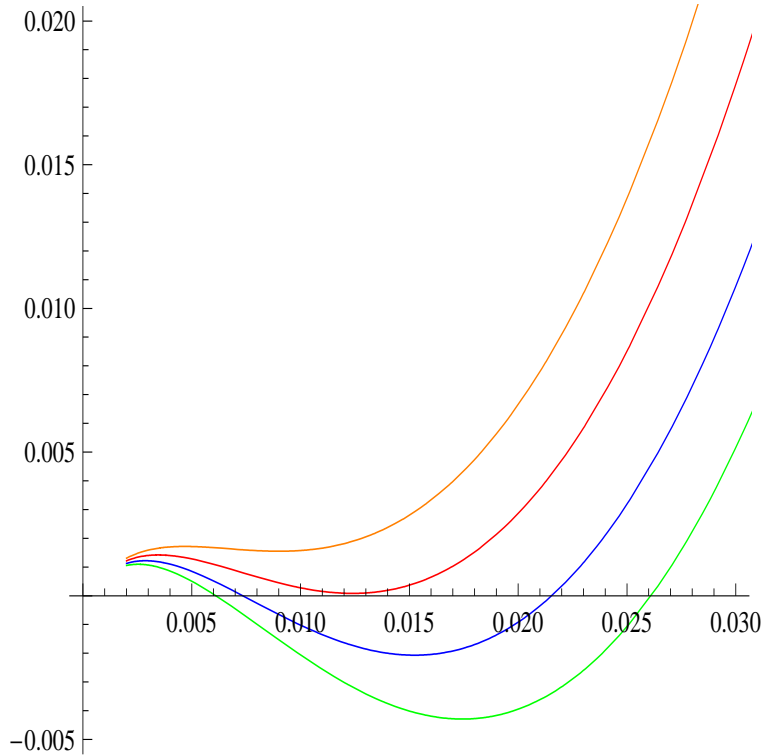}
\caption{$W_{BT}^{BI}$ is plotted against $Q$ using $x$ as a parameter for $n=4$. We have fixed $\phi=0.2$ and have plotted for different values of temperature. The red line is for $T=T_c$. The blue and green lines are for $T>T_c$ in an increasing order, whereas the orange line is for $T<T_c$ }
\label{bi2} 
\end{figure}

\noindent In order to study the phase structure, we use $x$, defined in eqn.(\ref{par}), as a parameter and carry out a parametric plot of $W_{BT}^{BI}$ against $Q$, the order parameter in the boundary theory. The resulting phase structure [fig.\ref{bi2}]\footnote{Again one expects the plots to go smoothly towards $Q=0$, which indeed is the case as can be checked from the free energy. But again by the same argument given in the previous footnote, this is not visible because of the choice of the plotting parameter.} shows a first order phase transition at some critical temperature, $T=T_c$ which turns out to be exactly the same as that in fig. \ref{bi}.\\\\
To conclude, for Reissner-Nordstr\"{o}m black hole, we are able to construct a candidate off-shell potential in terms of R-charge, $Q$, which, on-shell, gives all the stable phases of ${\cal{N}}= 4$ super Yang-Mills theory on $S^3$ at finite temperatures and finite non-zero chemical potentials. As for Born-Infeld black holes, an analytic construction becomes difficult. Via a semi-analytic approach, we showed that our construction leads to an effective potential with expected behaviour.\\\\
Now that we have a gauge theory effective potential, we could perhaps explore the details of the transition from the deconfining phase to the confining phase as we reduce the temperature.

\noindent 
\noindent{\large\bf{Acknowledgments:}}
First of all, I would like to thank Sudipta Mukherji for suggesting this problem to me, for extremely illuminating discussions at every step of the work and also for helping me in structuring and re-structuring the manuscript number of times. A very special thanks from me will always be there for Sayan Kumar Chakrabarti for innumerable fruitful discussion sessions and for taking immense pain in going through the manuscript and doing necessary corrections where required. I also thank Roberto Emparan for his comments on an earlier version of this paper. I am also grateful to Anirban Basu and Sachin Jain for many formal and informal discussions. Finally, I would like to thank Jnanadeva Maharana for several useful conversations and specially for providing me with encouragements all the time. 
\renewcommand{\thesection}{\Alph{section}} 
\setcounter{section}{0} 

\section{A Note on Scaling}
Our notion in this section is to consider the limit where the boundary of $AdS_{n+1}$ is $R^n$ (flat) instead of $R\times S^3$ (elliptical). For Reissner-Nordstr\"{o}m in an asymptotically AdS space in $(n+1)$ dimensions, the metric ansatz is similar to the Born-Infeld case, (\ref{2eq6}) and the solution thereof is 
\be
V(r) = k - \frac{m}{r^n-2} + \frac{q^2}{r^{2n-4}} + \frac{r^2}{l^2} ,
\ee  
\cite{emparan1} where k is related to scalar curvature. For elliptical horizon $k=1$, whereas for $k=0$, the horizon geometry will be flat. This solution can, infact, be obtained by taking $\beta\rightarrow\infty$ limit of (\ref{solmet}). Thus for $k=0$, 
\begin{equation}
ds^2 = -V(r)dt^2 + \frac{dr^2}{V(r)} + \frac{r^{2}}{l^2}\sum_{i=1}^{n-1}(dx_i)^2,
\label{rnf}
\end{equation}
with 
\be
V(r) = \frac{r^2}{l^2}- \frac{m}{r^{n-2}} + \frac{q^2}{r^{2n-4}}. 
\label{flatv}
\ee 
The limit in which one can go from the elliptic geometry of the horizon to a flat horizon is termed as `` infinite volume limit '', since the area of a flat horizon is infinite. This limit can be obtained by introducing a dimensionless parameter, $\lambda$ with which we scale different relevant quantities as \cite{emparan1}
\be
r\rightarrow\lambda^{\frac{1}{n}}r, t\rightarrow\lambda^{-\frac{1}{n}}t, m\rightarrow\lambda m, q\rightarrow\lambda^\frac{n-1}{n} q,
\ee 
and finally then taking $\lambda\rightarrow\infty$. Infact, one can check, this is precisely the limit in which $V(r)$ for $k=1$ reduces to that for $k=0$. Furthermore, the $(n-1)$ volume has also to be scaled as 
\be
l^2 d\Omega_{n-1}^2\rightarrow\lambda^{-\frac{2}{n}}\sum_{i=1}^{n-1}(dx_i)^2.
\label{scalevol}
\ee 
From (\ref{prn}), one can find the scaling for $\phi$,  
\be
\phi\rightarrow\lambda^{\frac{1}{n}}\phi.
\ee
In the same spirit, one can scale thermodynamic quantities too. Temperature, entropy, Energy and thermodynamic potential scale as \cite{son1}
\be
T\rightarrow\lambda^{\frac{1}{n}}T, S\rightarrow S,  E\rightarrow\lambda^{\frac{1}{n}}E, W\rightarrow\lambda^{\frac{1}{n}}W. 
\ee
The on-shell temperature, (\ref{temprn}), on rescaling and then taking $\lambda\rightarrow\infty$ limit, becomes
\be
T_{RN}\mid_{\lambda\rightarrow\infty}=\frac{n\rp^2-(n-2) c^2 l^2\phi^2}{4\pi l^2 \rp},
\label{tempinf}
\ee 
which is the same temperature as obtained directly by differentiating (\ref{flatv}) with respect to $\rp$ and dividing by $4\pi$ ( The Hawking temperature of a black hole, $T_H = \frac{\kappa}{2\pi}$ where $\kappa$ is the surface gravity given by $\kappa = -\frac{1}{2}\frac{\partial{g_{tt}}}{\partial{r}}\mid_{r=\rp}$. The physical reasoning behind this was discussed in section 3 in the context of Born-Infeld black holes. One can repeat the same with the $V(r)$ defined in (\ref{flatv}) and come across the same expression for temperature. )\\\\
For Reissner-Nordstr\"{o}m black holes in grand canonical ensemble, Energy, entropy and the Bragg-Williams free energy are given by
\bea
E&=& \frac{\omega_{n-1}}{16\pi G}(n-1)\Big[\rp^{n-2}(1+\phi^2 c^2)+\frac{\rp^n}{l^2}\Big],\\
S&=& \frac{\omega_{n-1}\rp^{n-1}}{4 G},\\
Q\phi &=& \frac{\omega_{n-1}}{8\pi G}\phi^2 c^2 (n-1)\rp^{n-2},\\
W_{BW}^{RN}&=&\frac{\omega_{n-1}}{16\pi G}\Big[(n-1)\rp^{n-2}(1-c^2{\phi}^2)-4\pi\rp^{n-1}T+\frac{\rp^n}{l^2}(n-1)\Big].
\eea
With the scaling defined above and taking the limit $\lambda\rightarrow\infty$ thereafter, those become
\bea
E&=& \lambda^{\frac{n-1}{n}}\frac{\omega_{n-1}}{16\pi G}(n-1)\Big[\rp^{n-2}\phi^2 c^2+\frac{\rp^n}{l^2}\Big],\\
S&=& \lambda^{\frac{n-1}{n}}\frac{\omega_{n-1}\rp^{n-1}}{4 G},\\
Q\phi &=& \lambda^{\frac{n-1}{n}}\frac{\omega_{n-1}}{8\pi G}\phi^2 c^2 (n-1)\rp^{n-2},\\
W_{BW}^{RN}&=&\lambda^{\frac{n-1}{n}}\frac{\omega_{n-1}}{16\pi G}\Big[(n-1)\frac{\rp^n}{l^2}-(n-1)\rp^{n-2}c^2{\phi}^2-4\pi\rp^{n-1}T\Big].
\eea
Thus on taking $\lambda\rightarrow\infty$ limit, all those quantities diverge. This is quite expected a result because, for a flat horizon geometry, the horizon area is infinity. So, instead of total energy, entropy and charge, one has to consider the corresponding densities. From (\ref{scalevol}), the $(n-1)$ volume ${\omega_{n-1}}$ should also scale as ${\omega_{n-1}}\rightarrow\lambda^{-\frac{n-1}{n}} \omega_{n-1}$. Then the energy density, entropy density and off-shell free energy density are given by 
\bea
\varepsilon &=& \frac{E}{\omega_{n-1}}= \frac{1}{16\pi G}(n-1)\Big[\rp^{n-2}\phi^2 c^2+\frac{\rp^n}{l^2}\Big],\\
s&=& \frac{S}{\omega_{n-1}} = \frac{\rp^{n-1}}{4 G},\\
\rho\phi &=& \frac{Q\phi}{\omega_{n-1}} = \frac{1}{8\pi G} \phi^2 c^2 (n-1)\rp^{n-2},\\
\Omega_{BW}^{RN}&=&\frac{W_{BW}^{RN}}{\omega_{n-1}}=\frac{1}{16\pi G}\Big[(n-1)\frac{\rp^n}{l^2}-(n-1)\rp^{n-2}c^2{\phi}^2-4\pi\rp^{n-1}T\Big].
\eea
$\frac{\partial{\Omega_{BW}^{RN}}}{\partial{\rp}}=0$ gives the correct on-shell temperature, (\ref{tempinf}).\\\\ 
Now following the discussion leading to eqn.(\ref{tcfc}) in section 6.1, one can check that there is no real solution for $T_c$ in this case. This is consistent with the infinite volume limit taken, because as we arrive at the flat horizon geometry, there will be only black hole phase and hence the possibility of Hawking-Page phase transition from black hole to AdS does not arise at all.


\begin{thebibliography}{99}

\bibitem{review} O. Aharony, S. Gubser, J. Maldacena, H.Ooguri and Y. Oz,
Phys. Rept. 323 (2000) 183, hep-th/9905111.

\bibitem{gkt} S. Gubser, I. Klebanov and A. Tseytlin, Nucl. Phys. {\bf 
B534} 
(1998) 202,
hep-th/9805156.

\bibitem{son1}
  D.~T.~Son and A.~O.~Starinets,
  JHEP {\bf 0603}, 052 (2006)
  [arXiv:hep-th/0601157].

\bibitem{cvetic}
  K.~Behrndt, M.~Cvetic and W.~A.~Sabra,
  Nucl.\ Phys.\  B {\bf 553}, 317 (1999)
  [arXiv:hep-th/9810227].

\bibitem{duff}
  M.~Cvetic {\it et al.},
  Nucl.\ Phys.\  B {\bf 558}, 96 (1999)
  [arXiv:hep-th/9903214].


\bibitem{emparan1} A. Chamblin, R. Emparan, C. Johnson and R. Myers, Phys.
Rev. {\bf D60} (1999) 064018, hep-th/9902170.

\bibitem{emparan2} A. Chamblin, R. Emparan, C. Johnson and R. Myers, Phys. 
Rev. {\bf D60} (1999) 104026, hep-th/9904197.


\bibitem{cvetic1}
  M.~Cvetic, S.~Nojiri and S.~D.~Odintsov,
  Nucl.\ Phys.\  B {\bf 628}, 295 (2002)
  [arXiv:hep-th/0112045].

\bibitem{nojiri}
  S.~Nojiri and S.~D.~Odintsov,
  Phys.\ Rev.\  D {\bf 66}, 044012 (2002)
  [arXiv:hep-th/0204112].

\bibitem{brigante}
  M.~Brigante, H.~Liu, R.~C.~Myers, S.~Shenker and S.~Yaida,
  Phys.\ Rev.\  D {\bf 77}, 126006 (2008)
  [arXiv:0712.0805 [hep-th]].

\bibitem{buchel}
  A.~Buchel, R.~C.~Myers and A.~Sinha,
  JHEP {\bf 0903}, 084 (2009)
  [arXiv:0812.2521 [hep-th]].

\bibitem{anindya}
  R.~C.~Myers, M.~F.~Paulos and A.~Sinha,
  JHEP {\bf 0906}, 006 (2009)
  [arXiv:0903.2834 [hep-th]].


\bibitem{krug}
  S.~Fernando and D.~Krug,
  Gen.\ Rel.\ Grav.\  {\bf 35}, 129 (2003)
  [arXiv:hep-th/0306120].

\bibitem{tanay}
  T.~K.~Dey,
  Phys.\ Lett.\  B {\bf 595}, 484 (2004)
  [arXiv:hep-th/0406169].

\bibitem{caisun}
  R.~G.~Cai and Y.~W.~Sun,
  JHEP {\bf 0809}, 115 (2008)
  [arXiv:0807.2377 [hep-th]].

\bibitem{tan}
  H.~S.~Tan,
  JHEP {\bf 0904}, 131 (2009)
  [arXiv:0903.3424 [hep-th]].



\bibitem{Fernando} S. Fernando, Phys.Rev.{\bf D74} (2006) 104032,
hep-th/0608040.


\bibitem{cai} R.G.Cai, D.W.Pang, A.Wang, Phys.Rev.{\bf D70} (2004) 124034, 
hep-th/0410158.


\bibitem{chaikin} P. Chaikin and T.C. Lubensky, Principles of Condensed Matter Physics, Cambridge Univ. Press, 
Chapter 4.


\bibitem{hartnoll}
  S.~A.~Hartnoll,
  Class.\ Quant.\ Grav.\  {\bf 26}, 224002 (2009)
  [arXiv:0903.3246 [hep-th]].


\bibitem{kobayashi}
  S.~Kobayashi, D.~Mateos, S.~Matsuura, R.~C.~Myers and R.~M.~Thomson,
  JHEP {\bf 0702}, 016 (2007)
  [arXiv:hep-th/0611099].





\bibitem{wald1}
  I.~Racz and R.~M.~Wald,
  Class.\ Quant.\ Grav.\  {\bf 9}, 2643 (1992).

\bibitem{witten} E. Witten, Adv.Theor.Math.Phys.{\bf2} (1998) 505-532, 
hep-th/9803131.


\bibitem{sudipta}
  S.~Jain, S.~Mukherji and S.~Mukhopadhyay,
  JHEP {\bf 0911}, 051 (2009)
  [arXiv:0906.5134 [hep-th]].





\end{thebibliography}
\end{document}